\documentstyle[11pt,epsf]{article}
\makeatletter
\newcommand{\artsectnumbering}{%
\@addtoreset{equation}{section}
\renewcommand{\theequation}{\thesection.\arabic{equation}}}
\makeatother  
\newcommand{\al}{\alpha}
\newcommand{\ap}{\simeq}
\newcommand{\bt}{\beta}
\newcommand{\bv}{\begin{verbatim}}
\newcommand{\de}{\delta}

\newcommand{\dt}{\dot}

\newcommand{\ev}{\end{verbatim}}
\newcommand{\fr}{\frac}

\newcommand{\ha}{\hat}
\newcommand{\hb}{\hbar}

\newcommand{\la}{\lambda}

\newcommand{\na}{\nabla}

\newcommand{\om}{\omega}
\newcommand{\ph}{\phi}
\newcommand{\ps}{\psi}
\newcommand{\rh}{\rho}

\newcommand{\si}{\sigma}
\newcommand{\ta}{\tau}

\newcommand{\Th}{\Theta}
\newcommand{\th}{\theta}
\newcommand{\vb}{\verb}

\newcommand{\be}{\begin{equation}}
\newcommand{\ee}{\end{equation}} %\indent}
\newcommand{\eei}{\end{equation}\indent\indent}
\newcommand{\bc}{\begin{center}}
\newcommand{\ec}{\end{center}}
\newcommand{\ber}{\begin{eqnarray}}
\newcommand{\ear}{\end{eqnarray}}
\newcommand{\ba}{\begin{array}}
\newcommand{\ea}{\end{array}}

\newcommand{\p}{\partial}

\def\case#1/#2{\textstyle\frac{#1}{#2} }
\begin{document}
%%%%%%%%%%%%%%%%%%%%%%%%%%%%%%%%%%%%%%%%%%%%%%%%
\title{Symmetry Breaking Using Fluids II:  Velocity Potential Method.}
\author{Mark D. Roberts, \\\\
Department of Mathematics and Applied Mathematics, \\ 
University of Cape Town,\\
Rondbosch 7701,\\
South Africa\\\\
roberts@gmunu.mth.uct.ac.za} 
\date{\today}
\maketitle
\vspace{0.1truein}
\bc Published:  Hadronic Journal {\bf 20}(1997)73-83.\ec
\bc Eprint: hep-th/9904079\ec
\bc Comments:  13 pages, no diagrams,  one table,  Latex2e.\ec
\bc 3 KEYWORDS:\ec
\bc Covariantly Charged Fluid:~~  Symmetry Breaking:~~  Higg's Model.\ec
\bc 1999 PACS Classification Scheme:\ec
\bc http://publish.aps.org/eprint/gateway/pacslist \ec
\bc 11.15Ex,  14.80Gt,05.70Jk,04.40+c\ec
\bc 1991 Mathematics Subject Classification:\ec
\bc http://www.ams.org/msc \ec
\bc 831T13,  83C55.\ec
\newpage
%%%%%%%%%%%%%%%%%%%%%%%%%%%%%%%%%%%%%%%%%%%%%%%%
\begin{abstract}
%%%%%%%%%%%%%%%%%%%%%%%%%%%%%%%%%%%%%%%%%%%%%%
A generalization of scalar electrodynamics called fluid 
electrodynamics is presented.  In this theory a fluid replaces the Higgs 
scalar field.   Fluid electrodynamics might have application to the theory 
of low temperature Helium superfluids,  but here it is argued that it provides 
an alternative method of approaching symmetry breaking in particle physics.   
The method of constructing fluid electrodynamics is to start with the 
velocity decomposition of a perfect fluid as in general relativity.   A unit 
vector tangent to the flow lines of an isentropic fluid can be written in 
terms of scalar potentials:  $V_a=h^{-1}(\ph_a+\al\bt_a-\th S)$.   
A novel interacting charged fluid can be obtained by applying the covariant 
derivative:  $D_a=\p_a+ieA_a$ to these scalar potentials.   
This fluid is no longer isentropic and there are choices 
for which it either obeys the second law of thermodynamics or not.   
A mass term of the correct sign occurs for the $A$ term in the stress,  
and this mass term depends on the potentials in the above vector.   
The charged fluid can be reduced to scalar electrodynamics and the 
standard approach to symmetry breaking applied;  alternatively a mass can be 
induced by the fluid by using just the thermodynamic potentials and then 
fixing at a critical point,  if this is taken to be the Bose condensation 
point then the induced mass is negligible.
\end{abstract}
%%%%%%%%%%%%%%%%%%%%%%%%%%%%%%%%%%%%%%%%%%%%%%%%%%%%%%%%%%%%%%%%%%%%%%%%%%%%
\section{Introduction.}
%%%%%%%%%%%%%%%%%%%%%%%%%%%%%%%%%%%%%%%%%%%%%%%%%%%%%%%%%%%%%%%%%%%%%%%%%%%%%%
In particle physics fields are given mass by assuming the existence of
Higgs scalar fields.   There are a number of unsatisfactory aspects to this 
procedure,  for example:   the Higgs scalar has not been experimentally found,
also it is required to have a mass term of the wrong sign.   An alternative 
procedure has been suggested,  Roberts (1989) \cite{bi:mdr89} in which fluids 
rather than scalar fields are responsible for vector fields acquiring mass.   
The motivation for using fluids is a principle \cite{bi:mdr89} which states 
that there is only one concept of mass in physics;  furthermore this concept 
of mass must be ultimately of gravitational origin.  The picture envisaged 
can be thought of as occurring in four steps or stages,  
these words have unwanted temporal connotations so that 
we refer instead to four basic ingredients.   The first ingredient is that the 
vector field is taken to exist,  and it is assumed to have a 
{\bf primitive stress}.   
The second ingredient is that the vector field has statistical 
properties which produce an {\bf effective fluid} that couples to the 
primitive stress.   
The third ingredient is that the coupling between the fluid and the 
primitive stress produces a {\bf mass term}.   The fourth ingredient 
is that the 
{\bf gravitational origin} of this is in some as yet undiscovered relationship 
between statistical mechanics and gravitational theory.   Those of a more 
prosaic disposition can simply regard fluids as an alternative technical means 
for inducing mass,  perhaps with application in the theory of superfluids,
Isreal (1981) \cite{bi:isreal}.
Scalar fields and fluids can be equated in several ways as shown for example 
in \cite{bi:mdr89}.   Usually fluids exhibit more freedom as they are 
parameterised by $p,  \mu, V_a$ however static scalar fields are equivalent 
to fluids with imaginary vector $V_a$ so that there are sometimes different 
qualitative properties,  an  example of this is the asymptotic properties,
Roberts (1998) \cite{bi:mdr98}of the space-time.   
The technical methods in \cite{bi:mdr98},  
uses what in section \ref{sec:II} is called the "already interacting" fluid.  
Here the technical procedure is to replace derivatives 
in the velocity decomposition of the fluid tangent vector $V_a$  by vector 
covariant derivatives.   The resulting "covariantly interacting"  fluid can 
be reduced to scalar electrodynamics, and then the standard symmetry breaking 
procedure applied.   The covariantly interacting fluid generalizes scalar 
electrodynamics and the extra degrees of freedom provide the scope for other 
symmetry breaking mechanisms to be used.   One mentioned here requires that 
only quantities of thermodynamic importance be retained in the mass breaking 
term,  these thermodynamic quantities can then be fixed at a critical value 
such as the Bose condensation value.   The drawback of this approach in its 
present form is that it necessitates the introduction of an ad hoc time 
interval;  here this is taken to be the Planck time.  It is found that the 
induced mass is negligible,  i.e. well beneath the experimental limit of 
$10^{-50}Kg.$ \cite{bi:GN}.   It is of interest to know if it could be 
demonstrated that the photon mass is exactly zero.  The Proca equation 
\cite{bi:IZ} p.135  necessitates  $m^2\na_aA^a=0$,  so that the existence 
of photon mass fixes the gauge.   The Bohm-Aharonov effect \cite{bi:BH} 
shows that the gauge must be chosen such that $A_a$ is continuous;  
it might be possible to devise a geometric configuration in which the 
continuity of $A_a$  requires a different gauge 
choice from $\na_aA^a=0$,  thus giving $m=0$ from $m^2\na_aA^a=0$.  
In this paper section \ref{sec:II} briefly introduces the standard 
scalar electrodynamic symmetry breaking,   
section \ref{sec:III} introduces the velocity decomposition for the 
velocity tangent to a fluid,  section \ref{sec:IV} 
applies the vector covariant derivative to this,  
section \ref{sec:V} mentions some possible ways in which 
the resulting fluid could break symmetry.   The conventions used are:  
signature $-+++$,  $";"$ signifies Christoffel covariant derivative,  
symmetrization is denoted by round brackets,  
e.g.  $T_{(ab)}=(T_{ab}+T_{ba})/2$,  anti-symmetrization is denoted 
by square brackets,  e.g.  $T_{[ab]}=(T_{ab}-T_{ba})/2$.
%%%%%%%%%%%%%%%%%%%%%%%%%%%%%%%%%%%%%%%%%%%%%%%%%%%%%%%%%%%%%%%%%%%%%%%%%%%
\section{Scalar Electrodynamics.}
\label{sec:II}
%%%%%%%%%%%%%%%%%%%%%%%%%%%%%%%%%%%%%%%%%%%%%%%%%%%%%%%%%%%%%%%%%%%%%%%%%%%%%%
The scalar electrodynamic Lagrangian \cite{bi:higgs},  
\cite{bi:IZ}p.68,  \cite{bi:HE}p.699 is
\be
L_{sel}=-D_a\ps D^a\bar{\ps}-V(\ps\bar{ps})-\fr{1}{4}F^2,
\label{eq:2.1}
\ee
where the covariant derivative is
\be
D_a\ps=\p_a\ps+ieA_a,
\label{eq:2.2}
\ee
and $D_a\bar{\ps}=\bar{D_a\ps}$.   
The variation of the corresponding action with respect to
$A_a,  \ps$, and $\bar{ps}$ are given by
\ber
\fr{\de I}{\de A_c}&=&F^{ab}_{..;b}
                     +ie(\ps D^a\bar{\ps}-\bar{ps}D^a\ps)\nonumber\\
\fr{\de I}{\de \ps}&=&(D_aD^a-V')\bar{\ps},~~~V'=\fr{dV}{d(\ps\bar{\ps})},
\label{eq:2.3}
\ear
and its complex conjugate.   Variations of the metric give the stress
\be
T_{ab}=2D_{(a}\ps D_{b)}\bar{\ps}+F_{ac}F^{~c}_{b.}+g_{ab}L.
\label{eq:2.4}
\ee
The complex scalar field can be put in "polar" form by defining
\be
\ps= exp(i\nu),
\label{eq:2.5}
\ee
giving the Lagrangian
\be
L=\rh^2_a+({\mathcal D}_a\nu)^2-V(\rh^2)-\fr{1}{4}F^2,
\label{eq:2.6}
\ee
where
\be
{\mathcal D}_a\nu=\rh(\nu_a+eA_a).
\label{eq:2.7}
\ee
The variations of the corresponding action with respect to 
$A_a   ,\rh$,  and $\nu$ are given by
\ber
\fr{\de I}{\de A_a}&=&F^{ab}_{..;b}+2e\rh{\mathcal D}^a\nu,\nonumber\\
\fr{\de I}{\de \rh}&=&2\left(\Box+(\nu_a+eA_a)^2+V'\right),\nonumber\\
\fr{\de I}{\de \nu}&=&2\left(\Box\nu+eA^a_{,a;a}\right).
\label{eq:2.8}
\ear
Variation of the metric gives the stress
\be
T_{ab}=2\rh_a\rh_b
      +2{\mathcal D}_a\nu{\mathcal D}_b\nu+F_{ac}F^{~c}_{b.} 
      +g_{ab}L.
\label{eq:2.9}
\ee
Defining
\be
B_a=A_a+\nu_a/e,
\label{eq:2.10}
\ee
$\nu$ is absorbed to give Lagrangian
\be
L=\rh^2_a+\rh^2e^2B_a^2-V(\rh^2)-\fr{1}{4}F^2,
\label{eq:2.11}
\ee
which does not contain $\nu$;  equation \ref{eq:2.10} 
is a gauge transformation when there are no discontinuities 
in $\nu$ i.e. $\nu_{;[ab]}=0$.

The requirement that the corresponding quantum theory is renormizable
restricts the potential to the form
\be
V(\rh^2)=m^2\rh^2+\la\rh^4.
\label{eq:2.12}
\ee
The ground state is when there is a minimum,  for $m^2,\la>0$ this is $\rh=0$,
but for $m^2<0,\la>0$ this is
\be
\rh^2=\fr{-m^2}{2\la}=a^2,
\label{eq:2.13}
\ee
thus the vacuum energy is
\be
<0|\rh|0>=a.
\label{eq:2.14}
\ee
To transform the Lagrangian \ref{eq:2.11} to take this into account substitute
\be
\rh\rightarrow\rh'=\rh+a,
\label{eq:2.15}
\ee
to give
\be
L=\rh_a^2+(\rh+a)^2e^2B_a^2-V\left((\rh+a)^2\right)-\fr{1}{4}F^2.
\label{eq:2.16}
\ee
Now apparently the vector field has a mass $m$ from the $a^2e^2B_a^2$ 
term it is given by $m=ae$.   The cross term $2a\rh e^2B_a^2$ is ignored.
%%%%%%%%%%%%%%%%%%%%%%%%%%%%%%%%%%%%%%%%%%%%%%%%%%%%%%%%%%%%%%%%%%%%%%%%%%%%
\section{Velocity Potentials.}
\label{sec:III}
%%%%%%%%%%%%%%%%%%%%%%%%%%%%%%%%%%%%%%%%%%%%%%%%%%%%%%%%%%%%%%%%%%%%%%%%%%%%
A Newtonian $3$-vector,  can be decomposed:  $\nu=\na\ph+\al\na\bt$,
\cite{bi:clebsch} where $\ph, \al$,  and $\bt$ are the Clebsch potentials.   
A particular case of Paff's theorem shows that a $4$-vector can be decomposed:
$V_a=AB_a+CD_a$ where $A,  B,  C$,  and $D$ are the potentials.   
The work of 
\cite{bi:lin}
\cite{bi:SW}
\cite{bi:schutz70}
\cite{bi:schmid1}
\cite{bi:schmid2}
shows that a non-minimal decomposition
\be
V_a=h^{-1}(\ph_a+\al\bt_a-\th S_a),~~~V_aV^a_.=-1
\label{eq:3.1}
\ee
is more useful,  because for an isentropic fluid all the potentials have 
evolution equations.   $\th$ is the thermasy of van Dantzig 
\cite{bi:vand}eq.4.9 it is usually defined by
\be
d\th=-kT~d\ta,
\label{eq:3.2}
\ee
where $T$ is the temperature;  also $h$ is the enthalpy and $S$ is the entropy.
The three other potentials do not have a thermodynamic interpretation.
A current vector can be defined by
\be
W_a=hV_a,
\label{eq:3.3}
\ee
c.f.\cite{bi:HE}p.69,  because $W^a_{.;a}=0$ it is not conserved.   
The Christoffel derivative of $W_a$ can be in the usual manner 
\cite{bi:HE}p.82-3
\be
W_{a;b}=\om_{ab}+\si_{ab}+\fr{1}{3}\Th h_{ab}-\dot{W}_aW_b,
\label{eq:3.4}
\ee
and then defining the vorticity tensor,  vorticity vector,  expansion tensor,
expansion scalar,  shear tensor,  and acceleration vector in the usual 
manner,   it is found that only the vorticity tensor, expansion scalar,  
and acceleration vector show any simplification,  they are
\ber
\om _{ab}&\equiv&h^{~c}_{a.}h^{~a}_{b.}W_{[c;d]}    
       =h^{~c}_{a.}h^{~a}_{b.}(\al_{[d}\bt_{c]}-\th_{[d}S_{c]}),\nonumber\\
\Th&\equiv&W^a_{.;a}
       =\Box\ph+\al\Box\bt-\th\Box S+\al^a\bt_a-\th^aS_a,\nonumber\\
\dot{W}_a&\equiv&V^bW_{a;b}
       =-TS_a+V^a(\ph_{ab}+\al\bt_{ab}-\th S_{ab}),
\label{eq:3.5}
\ear
respectively,  where the projection tensor is given by
\be
h_{ab}=g_{ab}-V_aV_b,
\label{eq:3.6}
\ee
and the evolution equations \ref{eq:3.13} have been used in deriving the 
equation for the acceleration $\dot{W}_a$.

The evolution equations can be derived from a Lagrangian;  to do this
it is necessary to assume both the equation
\be
nh=p+\mu,
\label{eq:3.7}
\ee
where $n$ is the particle number,  $p$ is the pressure,  and $\mu$ 
is the density,  and also the first law of thermodynamics in the form
\be
dp=n~dh-nT~dS.
\label{eq:3.8}
\ee
Usually \cite{bi:hargreaves} the the Lagrangian of a fluid is taken to be 
the pressure,  this is done here;  but occasionally. 
e.g.\cite{bi:HE}p.69, other quantities are used.  The action is
\be
I=\int\sqrt{-g}p~dx.
\label{eq:3.9}
\ee
Using \ref{eq:3.8} and then \ref{eq:3.1} shows that variations in the 
pressure depend on the velocity potentials.   The variations are
\ber
\fr{\de I}{\de g^{ab}}&=&-nhV_aV_b,\nonumber\\
\fr{\de I}{\de\ph}&=&-\sqrt{-g}(nV^a)_{;a},\nonumber\\
\fr{\de I}{\de\al}&=&-\sqrt{-g}n\bt^aV_a,\nonumber\\
\fr{\de I}{\de\bt}&=&-\sqrt{-g}(n\al V^a)_{;a},\nonumber\\
\fr{\de I}{\de\th}&=&+\sqrt{-g}nS^aV_a,\nonumber\\
\fr{\de I}{\de S}&=&-\sqrt{-g}[(n\th V^a)+nT].
\label{eq:3.10}
\ear
The first of these variations give the stress
\be
T_{ab}=(p+\mu)V_aV_b+pg_{ab}.
\label{eq:3.11}
\ee
The absolute derivative is given by
\be
\dot{X}_{ab\ldots}=\fr{D}{d\ta}X_{ab\ldots}=V^cX_{ab\ldots;c},
\label{eq:3.12}
\ee
then the requirement that the second of the variations 
\ref{eq:3.10} vanishes is
\be
\dot{n}+nV^a_{.;a}=0,
\label{eq:3.13}
\ee
and this is just the conservation of particle number \cite{bi:stewart}eq.2.3.
Using this the vanishing of the remaining variations gives
\be
\dot{\al}=\dot{\bt}=\dot{S}=0,~~~\dot{\th}= -T.
\label{eq:3.14}
\ee
The third of these shows that the fluid is isentropic,  the fourth is just 
\ref{eq:3.2} in another forms.   
Using \ref{eq:3.1} and \ref{eq:3.14} shows that
\be
\dot{\ph}=-h.
\label{eq:3.15}
\ee

The Bianchi identities give the fluid conservation equations
\ber
-V_aT^{ab}_{..;b}&=&\dot{\mu}+(p+\mu)V^a_{.;a},\nonumber\\      
h_{ab}T^{bc}_{..;c}&=&(p+\mu)V^bV_{a;b}+h^{~b}_{a.}p_{;b},
\label{eq:3.16}
\ear
where the projection tensor is given by \ref{eq:3.5};  these equations do not
immediately occur as a result of varying the Lagrangian,  but only by applying
the Bianchi identities to \ref{eq:3.11}.   The first law of thermodynamics 
has been assumed \ref{eq:3.5}.   In the present case the second law is obeyed 
as an equality as in \ref{eq:3.14} $\dot{S}=0$.   The situation is more 
complex in the next section.
Anticipating some of the problems and how to approach them -   In particle
physics there is not always invariance under time and space reflections.
The above involves no inequalities:  there are no assumed energy inequalities,
and the fluid is isentropic (in equilibrium with $\dot{S}=0$) rather than
$\dot{S}>0$.   Recall that a vector is future pointing iff $V_t>0$;  
for \ref{eq:3.1} $hV_t=\ph_t+\al\bt_t-\th S_t$;  ignoring $\ph,  \al$,   
and $\bt$,   $V_a$ is past pointing when  $\th S_t >0$.   
This can be reversed by either defining a new vector $V'_a=-V_a$ or 
a new time coordinate $t'=-t$.
%%%%%%%%%%%%%%%%%%%%%%%%%%%%%%%%%%%%%%%%%%%%%%%%%%%%%%%%%%%%%%%%%%%%%%%%%%%%
\section{The "Covariantly Interacting" Fluid.}
\label{sec:IV}
%%%%%%%%%%%%%%%%%%%%%%%%%%%%%%%%%%%%%%%%%%%%%%%%%%%%%%%%%%%%%%%%%%%%%%%%%%%%%%%
There are several ways,  four of which are mentioned here,  of introducing
an interacting fluid and vector field,  with the intention of breaking the 
vector fields symmetries.   In the first method a "plasma interacting" fluid
is produced by generalizing the treatment of \cite{bi:SW}$\vb+#+$7.   
In the second method,   the fluid is "already interacting",  
the stress \ref{eq:3.10} is directly equated
with the stress calculated from \ref{eq:2.11}; 
this gives $-2(\rh_a\rh_b+\rh^2e^2B_aB_b)=(\mu+p)V_aV_b$ 
and it is impossible to proceed with one real fluid;  this method is similar 
to the method discussed in \cite{bi:mdr89}.   
The third method is the "traditionally 
interacting" fluid \cite{bi:HE}p.70,  this is produced by adding a term
$L_I=-\fr{1}{2}V_aA^a$ to the Lagrangian;  
this method is of no use for present purposes as there is
only a single term $A_a$ in the Lagrangian and stress, 
and for symmetry breaking products $A_a A_B$ are required.   
The fourth method is the "covariantly interacting"  fluid;   
this is a produced by simply applying the covariant 
derivative \ref{eq:2.2} to the vector \ref{eq:3.1} to produce
\be
V_a=h^{-1}\left(\ph_a+\al\bt_a-\th S_a+ie(\ph+\al\bt-\th S)A_a\right).
\label{eq:4.1}
\ee
It is required that
\be
-1=g^{ab}_{..}V_{(a}\bar{V}_{b)},
\label{eq:4.2}
\ee
and also that the projection tensor is
\be
h_{ab}=g_{ab}-V_{(a}\bar{V}_{b)},
\label{eq:4.3}
\ee
this ensures that most quantities are real.   After using $W_a=hV_a$,
\ref{eq:4.2} becomes
\be
h=-g^{ab}_{..}W_{(a}\bar{V}_{b)},
\label{eq:4.4}
\ee
then via the first law of thermodynamics \ref{eq:3.7} the pressure and 
the Lagrangian are real;  however the expansion,  shear,  \ldots etc. 
are complex,  this can be verified by direct computation or by noting 
from \ref{eq:3.4} that as $W_a$ is complex the quantities involved 
in this decomposition should also be complex.

The evolution equations are derived in a similar manner to the derivation 
in the last section.   The action is taken to be \ref{eq:3.9} 
and the first law of thermodynamics is assumed with the enthalpy $h$ 
now given by \ref{eq:4.4}.   The variations of the action are
\ber
\fr{\de I}{\de g^{ab}_{..}}&=&-nhV_{(a}\bar{V}_{b)},\nonumber\\
\fr{\de I}{\de A^a_.}&=&-n\sqrt{-g}\fr{e'^2}{h^2}A_a,
~~~e'^2=e^2(\ph+\al\bt-\th S),\nonumber\\
\fr{\de I}{\de \ph}&=&-\sqrt{-g}[(n\Re(V^a_.))+e'^2A_a^2n],\nonumber\\
\fr{\de I}{\de \al}&=&-n\sqrt{-g}[\bt_a\Re(V^a_.)+e'^2A_a^2\bt],\nonumber\\ 
\fr{\de I}{\de \bt}&=&-\sqrt{-g}[(n\al\Re(V^a_.)_{;a}
                                              +e'^2A_a^2n\al],\nonumber\\ 
\fr{\de I}{\de \th}&=&+n\sqrt{-g}[S_a\Re(V^a_.)+e'^2A_a^2S],\nonumber\\
\fr{\de I}{\de   S}&=&+\sqrt{-g}[(n\th\Re(V^a_.))_{;a}+e'^2A_a^2\th n-nT],
\label{eq:4.5}
\ear
\ref{eq:4.4} and the vanishing of these give
\be
-h=[\ph_a\Re(V^a_.)+e'^2A_a^2\ph].
\label{eq:4.6}
\ee
The stress and Maxwell equation are
\ber
T_{ab}&=&(p+\mu)V_{(a}\bar{V}_{b)}+p~g_{ab},\nonumber\\
F^{ab}_{..}&+&\fr{e}{h^2}(\ph+\al\bt-\th S)A^a_.= 0.
\label{eq:4.7}
\ear
Introducing the notation
\ber
\dt{X}_{ab\ldots}&=&\fr{DX_{ab\ldots}}{d\ta}
                   =\Re(V^c_.)X_{ab\ldots;c},\nonumber\\
\ha{X}_{ab\ldots}&=&\left(\fr{D}{d\ta}+e'^2A_a^2\right)X_{ab\ldots},\nonumber\\
\label{eq:4.8}
\ear
\ref{eq:4.6} and the vanishing of the variations \ref{eq:4.5} 
can be written in the simple form
\be
\dt{\al}=\ha{\bt}=\ha{S}=0,~~~\ha{\ph}=-h,~~~\dt{\th}=T,~~~\ha{n}=n\Th\ne0.
\label{eq:4.9}
\ee
Changing the sign and index of the last term produces the changes in the table.
The fluid has $n\ne=0$ implying that particle number is not conserved,  this 
appears unavoidable.   Our choice has $\th=T$ so that the standard expression 
for thermasy is recovered;  also $\ha{S}=0$ implies that the second law of 
thermodynamics is obeyed when $\th S>\ph+\al\bt$.\newpage
Table:  The Choice for the Thermodynamic Term in the Velocity Vector.
\begin{tabular}{||l|l|l|l||}                               \hline\hline
$  -\th_a S $    & $ +\th S_a $  &  $ +\th_a S  $   & $ -\th S_a$\\   \hline
$\ha{\ph}=ST-h$  & $ \ha{\ph}=-h$&  $\ha{\ph}=ST-h$ & $\ha{\ph}=-h$\\
$ \dt{S}=0 $     & $ \ha{S}=0 $  &  $\dt{S}=0   $   & $\ha{S}=0$\\ 
$ \ha{\th}=-T $  & $ \dt{\th}=-T$&  $\ha{\th}=-T$   & $\dt{\ph}=-T$\\ \hline\hline
\end{tabular}
%%%%%%%%%%%%%%%%%%%%%%%%%%%%%%%%%%%%%%%%%%%%%%%%%%%%%%%%%%%%%%%%%%%%%%%%%%%%
\section{Symmetry Breaking.}
\label{sec:V}
%%%%%%%%%%%%%%%%%%%%%%%%%%%%%%%%%%%%%%%%%%%%%%%%%%%%%%%%%%%%%%%%%%%%%%%%%%%%
The straightforward way of proceeding to break symmetry is to note that 
the covariantly interacting fluid of the previous section contains scalar 
electrodynamics as a special case.   There are several choices of the 
parameters by which scalar electrodynamics can be recovered,  an example is
\be
\al=\bt=\th=S=\p_a h=0,~~~\ph=\sqrt{2}h\ps,~~~p=l_{sel},~~~\mu=1-p,
\label{eq:5.1}
\ee
in the gauge ref{eq:2.10}.   Thus the standard way of breaking symmetry 
can then be applied,  with the aesthetic difference the fundamental 
cause of breaking is a fluid not a field.   
There is the possibility of other velocity potentials such as $\al$    
being non-vanishing operators leading to modifications of the standard 
treatment;   the velocity potentials themselves have been subject
to quantization for a fluid coupled to gravity in the ADM formalism,  see
for example 
\cite{bi:schutz71}
\cite{bi:DM}
\cite{bi:LR}.

The Proca equation \cite{bi:IZ}p.135 is
\be
F^{ab}_{..}+m^2eA^a_.=0,
\label{eq:5.2}
\ee
comparing with (\ref{eq:4.8},  the covariantly interacting fluid has a mass
\be
m^2=h^{-2}(\ph+\al\bt-\th S)^2.
\label{eq:5.3}
\ee
Choosing only to retain thermodynamic quantities $m^2=h^{-2}\th^2S^2$, and the 
thermasy $\th$ needs to be evaluated.   The tempreture can be taken to be 
independent of the proper time $\ta$,  but the mass in \ref{eq:5.3} 
still depends on proper time;  to proceed it is necessary to introduce 
an artificial time interval and the Plank time is chosen,  thus
\be
-\th= k\int^{ta_{PL}}_0~T~d\ta
    = kT\int^{\ta_{PL}}_0~d\ta
    =kT~|^{\ta_{PL}}_0
    =\sqrt{\fr{G\hb}{c^5}}kT.
\label{eq:5.4}
\ee
Restoring constants and substituting into \ref{eq:5.3} gives
\be
m=\sqrt{\fr{c\hb}{G}}k\fr{TS}{h}\ap3.10^{-31}\fr{TS}{h}~ Kg.
\label{eq:5.5}
\ee
For $m$ to be constant it is necessary to evaluate $TS/h$ at the critical point
and the Bose condensation point is a choice.   This choice terms out to give
zero mass because Bose condensation requires that the lowest state has zero
kinetic energy,  as there is no extra energy there is no extra heat content
and hence no entropy,  there is no disorder because all the particles are in 
the same state,  c.f.\cite{bi:kahanna}p.78,  
no entropy implies that there is no induced mass.   
The Bose condensation point is also unsatisfactory choice because 
strictly speaking $N/V$ is not required to have a given value for radiation in
a cavity for a vector fields,  c.f.\cite{bi:jackson}prob.16.2.
%%%%%%%%%%%%%%%%%%%%%%%%%%%%%%%%%%%%%%%%%%%%%%%%%%%%%%%%%%%%%%%%%%%%%%%%%%%%%%


\begin{thebibliography}{22}
%%%%%%%%%%%%%%%%%%%%%%%%%%%%%%%%%%%%%%%%%%%%%%%%%%%%%%%%%%%%%%%%%%%%%%%%%%%%%%
\bibitem{bi:mdr89}
M.D.Roberts,  
{\it Hadronic J.}{\bf12}(1989)93.

\bibitem{bi:mdr98}
M.D.Roberts,
Space-time Exterior to a Star,  gr-qc/9811093

\bibitem{bi:GN}
A.S.Goldberger and M.M.Nieto,  
{\it Rev.Mod.Phys.}{\bf 43}(1971)277.

\bibitem{bi:IZ}
C.Itzykson and J.-B.Zuber,\\  
Quantum Field Theory,\\  
McGraw-Hill (1985).

\bibitem{bi:BH}
D.Bohm and B.Hiley,  
{\it Nuovo Cim.}{\bf A52}(1979)295.

\bibitem{bi:higgs}
P.W.Higgs,  
{\it Phys.Lett.}{\bf 12}(1964)132.
 
\bibitem{bi:HE}
S.W.,Hawking and G.F.R.Ellis,\\  
The Large-Scale Structure of Space-time,\\
Cambridge University Press (1973).

\bibitem{bi:clebsch}
A.Clebsch,
{\it J.Riene Agnew Math.}{\bf 56}(1859)1.

\bibitem{bi:lin}
C.C.Lin,
in Liquid Helium,\\  
International School of Physics,  "Enrico Fermi",21(1963),\\ 
ed. G.Careri,  Academic Press, NY(1963).

\bibitem{bi:SW}
R.L.Seliger and G.B.Whitham, 
{\it Proc.Roy.Soc.Lond.}{\bf A305}(1968)1.

\bibitem{bi:schutz70}
B.F.Schutz,  
{\it Phys.Rev.}{\bf D2}(1970)2762.

\bibitem{bi:schmid1}
L.A.Schmid,
{\it Pure and Applied Chem.}{\bf 22}(1970)493.

\bibitem{bi:schmid2}
L.A.Schmid, Effects of Heat Exchange on Relativistic Fluid Flows,\\
in a Critical Review of Thermodynamics,\\  
eds.B.Gal-Or and A.J.Brainard, pp.161-202,  Mono,  Baltimore, (1970).

\bibitem{bi:vand}
D. van Dantzig, 
{\it Physica}{\bf 6}(1939)673.

\bibitem{bi:hargreaves}
R.Hargreaves, 
{\it Phil.Mag.}{\bf 16}(1908)436.

\bibitem{bi:stewart}
J.M.Stewart,  
{\it Proc.Roy.Soc.London.}{\bf A357}(1977)59.

\bibitem{bi:schutz71}
B.F.Schutz,
{\it Phys.Rev.}{\bf D4}(1971)3559.

\bibitem{bi:DM}
J.Demaret and V.Moncrief,
{\it Phys.Rev.}{\bf D21}(1980)2785.

\bibitem{bi:LR}
V.G.Lapchinskii and V.A.Rubakov,
{\it Theor.Math.Phys.}{\bf 33}(1977)1076.

\bibitem{bi:kahanna}
K.M.Kahanna,\\ 
Statistical Mechanics,\\  
Asia Publishing House,  London (1968).

\bibitem{bi:jackson}
K.M.Jackson,\\  
Equilibrium Statistical Mechanics,\\  
Prentice Hall, N.J.(1968).

\bibitem{bi:isreal}
W.Isreal,  {\it Phys.Lett.}{\bf A86}(1981)79.

\end{thebibliography}
\end{document}